# Lightweight Mutual Authentication Protocol for Low Cost RFID Tags

Eslam Gamal Ahmed[1], Eman Shaaban[2], Mohamed Hashem[3]

Faculty of Computer and Information Science
Ain Shams University
Abbasiaa, Cairo
EGYPT
{Eslam_Gamal[1], Eman.Shaaban[2], Mhashem[3]}@cis.asu.edu.eg

## ABSTRACT

*Radio Frequency Identification (RFID) technology one of the most promising technologies in the field of ubiquitous computing. Indeed, RFID technology may well replace barcode technology. Although it offers many advantages over other identification systems, there are also associated security risks that are not easy to be addressed. When designing a real lightweight authentication protocol for low cost RFID tags, a number of challenges arise due to the extremely limited computational, storage and communication abilities of Low-cost RFID tags.*

*This paper proposes a real mutual authentication protocol for low cost RFID tags. The proposed protocol prevents passive attacks as active attacks are discounted when designing a protocol to meet the requirements of low cost RFID tags. However the implementation of the protocol meets the limited abilities of low cost RFID tags.*

## 1 Introduction

Radio Frequency Identification (RFID) system is the latest technology that plays an important role for object identification as ubiquitous infrastructure. RFID has many applications in access control, manufacturing automation, maintenance, supply chain management, parking garage management, automatic payment, tracking, and inventory control.

EPCglobal is a member-driven organization composed of leading firms and industries that are focused on creating global standards for the EPCglobal Network. EPCglobal is now leading the development of industry-driven standards for the Electronic Product Code (EPC) Network to support the use of Radio Frequency Identification (RFID) in today's fast-moving, information rich trading networks [13].

An RFID system consists of three different components: RFID tag or transponder, Reader or interrogator, and backend server.

RFID tag: is a tiny radio chip that comprises a simple silicon microchip attached to a small flat aerial and mounted on a substrate. The whole device can then be encapsulated in different materials (such as plastic) dependent upon its intended usage. The tag can be attached to an object, typically an item, box, or pallet, and read remotely to ascertain its identity, position, or state. For an active tag there will also be a battery.

Reader or Interrogator: sends and receives RF data to and from the tag via antennas. A reader may have multiple antennas that are responsible for sending and receiving radio waves.

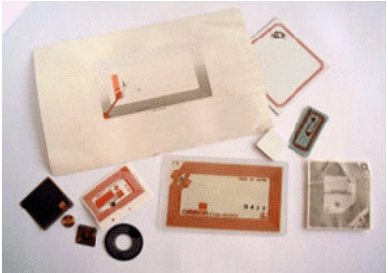
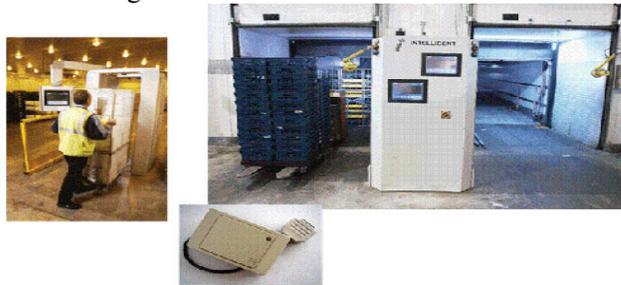

Figure 1. A variety of RFID Tags     Figure 2. Examples of a Reader with Associated Electronics

Host Computer (backend server): the data acquired by the readers is then passed to a host computer, which may run specialist RFID software or middleware to filter the data and route it to the correct application, to be processed into useful information.

RFID offer several advantages over barcodes: data are read automatically, line of sight not required, and through non-conducting materials at high rate and far distance. The reader can read the contents of the tags by broadcasting RF signals via antennas. The tags data acquired by the readers is then passed to a host computer, which may run middleware (API). Middleware offers processing modules or services to reduce load and network traffic within the back-end systems. RFID basic operations can be summarized as in Figure 3.

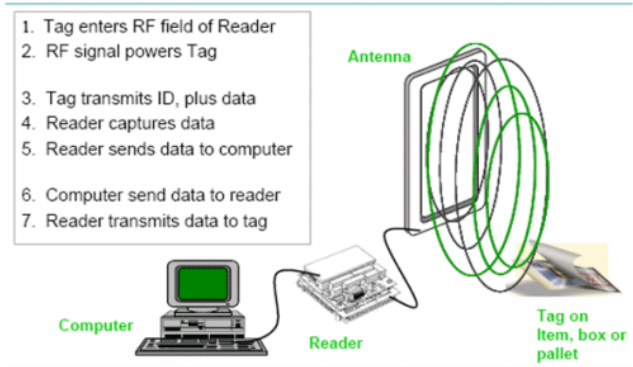

Figure 3. Basic Operations of RFID

RFID systems are vulnerable to a broad range of malicious attacks ranging from passive eavesdropping to active interference. Unlike in wired networks, where computing systems typically have both centralized and host-based defenses (e.g. firewalls), attacks against RFID networks can target decentralized parts of the system infrastructure, since RFID readers and RFID tags operate in an inherently unstable and potentially noisy environment. Additionally, RFID technology is evolving quickly – the tags are multiplying and shrinking - and so the threats they are susceptible to, are similarly evolving. Thus, it becomes increasingly difficult to have a global view of the problem [1].

RFID tags may pose a considerable security and privacy risk to organizations and individuals using them. Since a typical tag answers its ID to any reader and the replied ID is always the same, an attacker can easily hack the system by reading out the data of a tag and duplicating it to bogus tags. Unprotected tags may have vulnerabilities to eavesdropping, location privacy, spoofing, or denial of service (DoS). Unauthorized readers may compromise privacy by accessing tags without adequate access control. Even when the content of the tags is protected, individuals may be tracked through predictable tag responses. Even though many cryptographic primitives can be used to remove these vulnerabilities, they cannot be applied to a RFID system due to the prohibitive cost of including protection for each and every RFID tag [3].

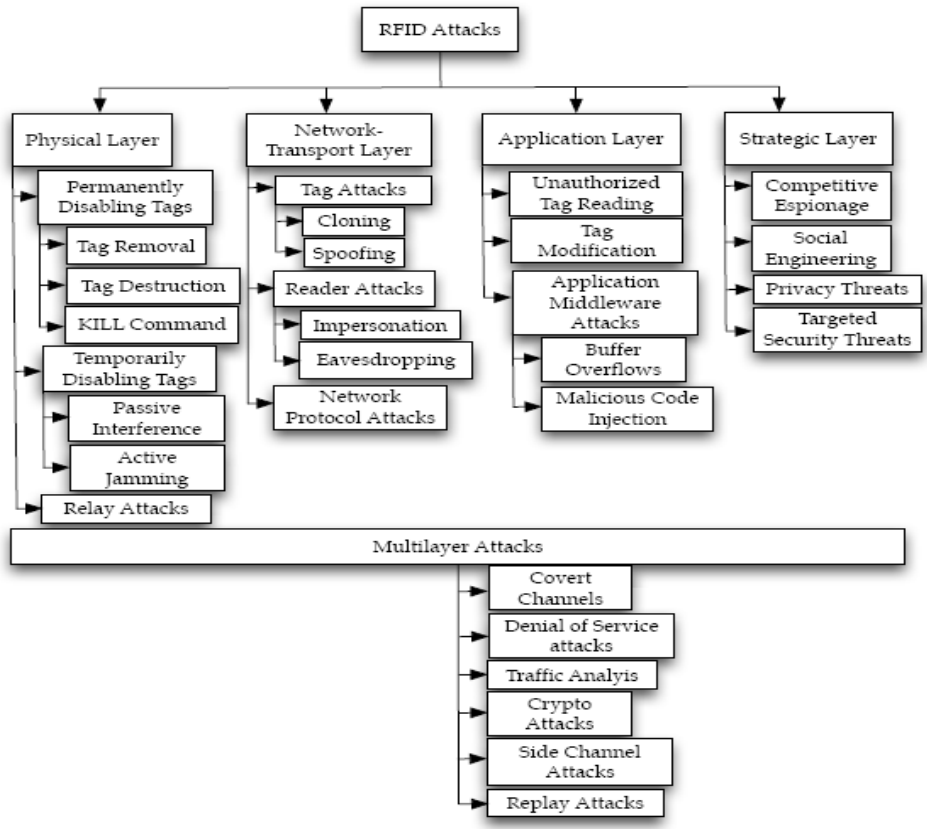

Figure 4. Classification of RFID attacks

Low-cost RFID tags in the form of Electronic Product Codes (EPC) are poised to become the most pervasive device in history. Already, there are billions of RFID tags on the market, used for applications like supply-chain management, inventory monitoring, access control, and payment systems. When designing a real lightweight authentication protocol for low cost RFID tags, a number of challenges arise due to the extremely limited computational, storage and communication abilities of low-cost RFID tags [4, 5, 7].

This paper proposes a real mutual authentication protocol for low cost RFID tags. The proposed protocol prevents passive attacks as active attacks are discounted when designing a protocol to meet the requirements of low cost RFID tags. However the implementation of the protocol meets the limited abilities of low cost RFID tags.

This paper is organized in the following way: Section 2 presents the related work for mutual authentication protocols in low cost RFID tags. Section 3 presents the analysis of Gossamer protocol. Section **4 presents a Passive attack against Gossamer.** Section 5 presents the proposed protocol. Section 6 presents the analysis of the proposed protocol. Section 7 concludes the paper.

## 2  Related Work

Over the past few years, researchers have developed many Ultralightweight Mutual Authentication Protocols (UMAP) that claim to prevent vulnerabilities associated with secure authentication of RFID tags and readers. In this section we consider a few recently proposed protocols and identify possible vulnerabilities in them.

The security of UMAP has been analyzed in depth by the research community. Researches show how a passive attacker can disclose part of the secret information stored in the tag's memory. Active attacks are discounted when designing a protocol to meet the requirements of ultralightweight RFID tags.

In 2006, Peris et al. proposed a family of Ultralightweight Mutual Authentication Protocols (henceforth referred to as the UMAP family of protocols). Chronologically, M2AP [12] was the first proposal, followed by EMAP [10] and LMAP [11].

In 2007 Chien proposed a very interesting lightweight authentication protocol providing Strong Authentication and Strong Integrity (SASI) for low-cost RFID tags [4]. An index-pseudonym (IDS), the tag's private identification (ID), and two keys (k1/k2) are stored both on the tag and in the back-end database. Simple bitwise XOR ($\oplus$), bitwise AND ($\wedge$), bitwise OR ($\vee$), addition $2^m$ and left rotate Rot(x,y) are required on the tag. Additionally, random number generation is required on the reader. The protocol is divided into three states: tag identification, mutual authentication and updating phase. In the identification phase, the reader (R) sends a "hello" message to the tag (T), and the tag answers with its IDS. The reader then finds, in the back-end database, the information associated with the tag (ID and k1/k2), and the protocol continues to the mutual authentication phase. In this phase, the reader and the tag authenticate each other, and the index-pseudonym and keys are subsequently updated.

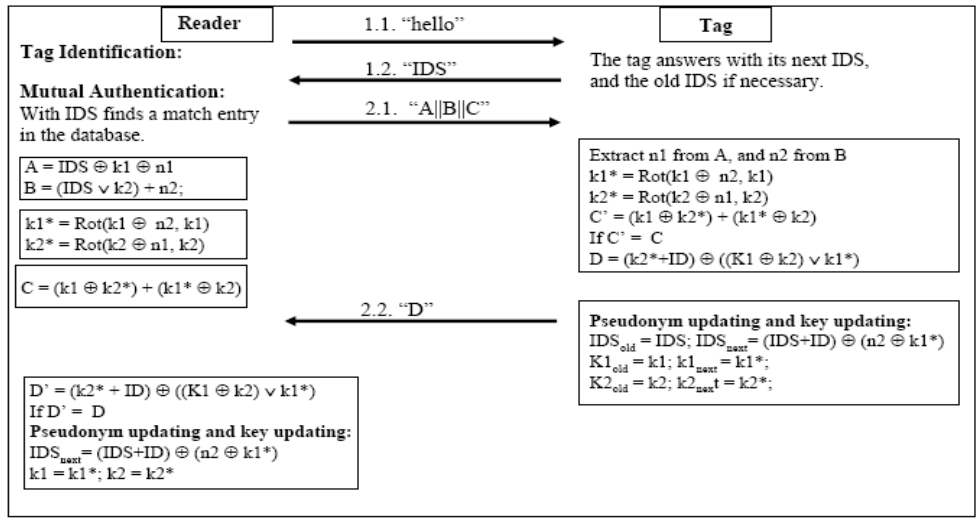

Figure 5. SASI Protocol

The protocol SASI was a step further towards a secure protocol compliant with real ultralightweight tag requirements. However cryptanalysis of SASI showed that passive attacker can obtain the secret ID by observing several consecutive authentications sessions [2].

**Cryptanalysis of SASI protocol:**

Assume the amount of rotation given by the second argument is zero modulo 96 we therefore have:
K1' = Rot (k1 $\oplus$ n2, k1) = Rot (k1 $\oplus$ n2, k1 mod 96) = Rot (k1 $\oplus$ n2, 0) = (k1 $\oplus$ n2).
Similarly,
k2' = Rot (K2 $\oplus$ n1, K2) = (K2 $\oplus$ n1).
So $IDS^{next}$ = (IDS + ID) $\oplus$ (n2 $\oplus$ k1') = (IDS + ID) $\oplus$ (n2 $\oplus$ k1 $\oplus$ n2) = (IDS + ID) $\oplus$ k1
So ID = ($IDS^{next}$ $\oplus$ k1−IDS)
and we can take full advantage of the knowledge that k1 = k2 = 0 mod 96 to conclude that:-
**ID mod 96 $\approx$ ($IDS^{next}$ − IDS) mod 96.            (1)**
AS C = (k1 $\oplus$ k2') + (k2 $\oplus$ k1') = (k1 $\oplus$ K2 $\oplus$ n1) + (K2 $\oplus$ k1 $\oplus$ n2)
This implies that
C mod 96 = (k1 $\oplus$ K2 $\oplus$ n1) + (K2 $\oplus$ k1 $\oplus$ n2) mod 96 $\approx$ (n1 + n2) mod 96.
The value of n1 +n2 mod 96 can also be obtained from the observed values of public messages A, B and IDS because
  A = (IDS $\oplus$ k1 $\oplus$ n1)           n1 ➔ (A $\oplus$ IDS $\oplus$ k1)
And then we can get that
n1 mod 96 = (A$\oplus$IDS$\oplus$ k1) mod 96 $\approx$ (A $\oplus$ IDS) mod 96

Similarly, $n2 \approx (B - IDS) \mod 96$
So $C \mod 96 \approx (n1 + n2) \mod 96$

$$C \mod 96 \approx (A \oplus IDS) + (B - IDS) \mod 96 \qquad (2)$$

Then snoop on multiple authentication sessions and, for each one, verify if previous condition holds. If this is the case, compute the value $(IDS^{next} - IDS) \mod 96$ and, from this, directly approximate $ID \mod 96$ **(1)**.

Hence Gossamer protocol which is inspired by SASI scheme but hopefully avoiding its weaknesses was proposed. The protocol comprises three stages: tag identification phase, mutual authentication phase, and updating phase Figure.6.

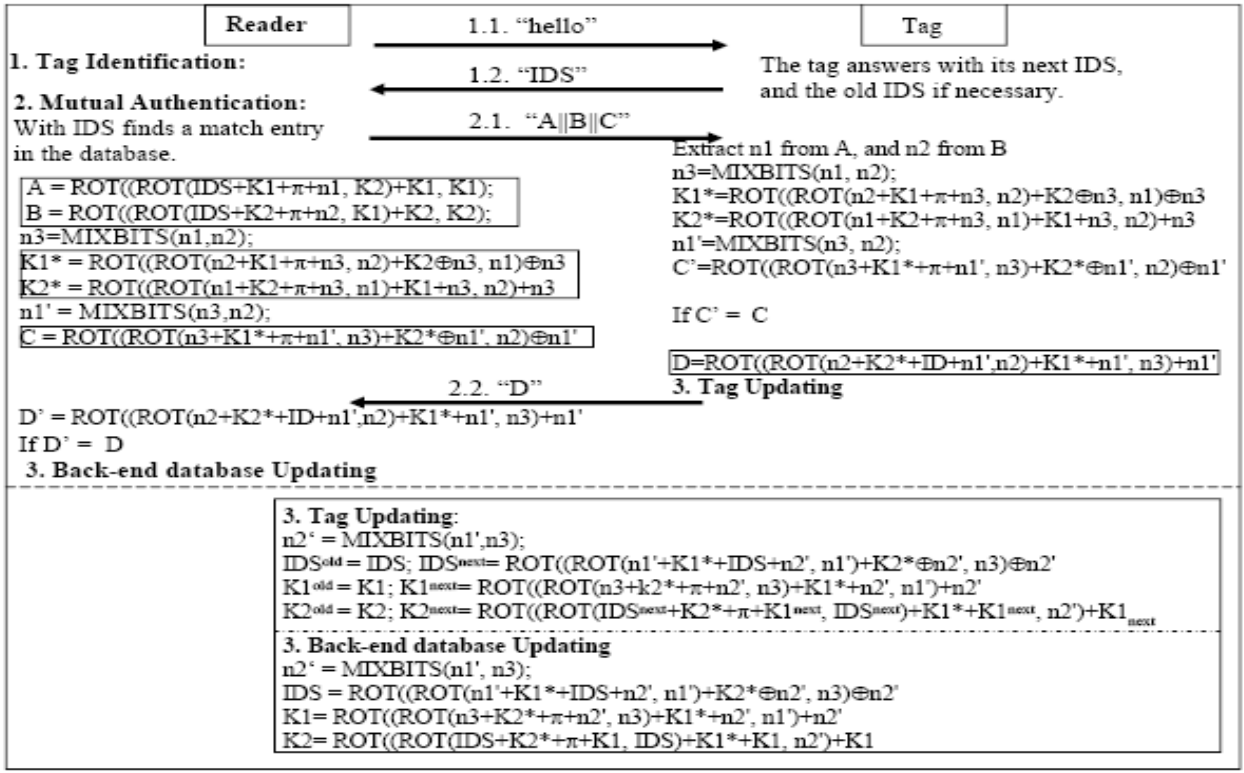

Figure 6. Gossamer Protocol

In Tag Identification the reader first sends a "hello" message to the tag, which answers with its potential next IDS. The reader then tries to match each entry in the database. If this search succeeds, the mutual authentication phase starts. Otherwise the identification is retried but with the old IDS, which is backscattered by the tag upon request. Mutual Authentication with IDS, the reader acquires the private information linked to the tag, identified from the database. Then the reader generates nonces n1 and n2 and builds and sends to the tag A||B||C. From submessages A and B, the tag extracts nonces n1 and n2. Then it computes n3/n'1 and k*1/k*2 and builds a local version of submessage C'. This is compared with the received value. If it is verified, the reader is authenticated. Finally, the tag sends message D to the reader. On receiving D, this value is compared with a computed local version. If comparison is successful, the tag is authenticated; otherwise the protocol is abandoned. Index-Pseudonym and Key Updating After successfully completing the mutual authentication phase between reader and tag, they locally update IDS and keys (k1/k2). Submessages C/D allow reader/tag authentication, respectively. Moreover, the use of submessages C/D results in confirmation of synchronization for the internal secret values (n3/n'1 and k*1/k*2) used in the updating phase, preventing straightforward desynchronization attacks. For the implementation of the proposed protocol, only simple operations are available on tags, in accordance with their restrictions: specifically, bitwise XOR (⊻), addition mod 2m (+), and Rot(x, y). To avoid ambiguity, Rot(x, y) is defined to perform a circular shift on the value of x, (y mod N) positions to the left for a given value of N (in our case 96). Random number generation, required in the protocol to supply freshness, is a costly operation, so it is performed by the reader. To significantly increase security, a non linear lightweight function calledMixBits is added:

$Z = MixBits(X, Y)$
{Z = X;
for (i=0; i<32; i++) {
Z = (Z>>1) + Z + Z + Y ;}}

## 3  Analysis of Gossamer protocol

This section presents the performance and security analysis of Gossmer protocol as presented in authors' papers [2].

### 3.1  Performance Analysis

**Computational cost:** Gossamer only requires simple bitwise XOR, addition $2^m$, left rotation, and the MixBits function on tags. These operations are very low-cost and can be efficiently implemented in hardware. MixBits is very efficient from a hardware perspective. The number of iterations of this function is optimized to guarantee a good diffusion effect.

**Storage requirement:** Each tag stores its static identifier (ID) and two records of the tuple (IDS, k1, K2) with old and potential new values. A 96-bit length is assumed for all elements in accordance with EPCGlobal. The ID is a static value, thus stored in ROM. The remaining values (96 × 6 = 576 bits) are stored in a rewritable memory because they need to be updated.

**Communication cost:** Gossamer performs mutual authentication and integrity protection with only four messages. In the identification phase, a "hello" and IDS message are sent over the channel. Messages A||B||C and D are transmitted in the authentication phase. So a total of 424 bits are sent over the channel - considering 5 bytes for the "hello"message.

### 3.2  Security Analysis

**Data Confidentiality** All public messages are composed of at least three secret values shared only by legitimate readers and genuine tags. The static identifier and the secret keys cannot, be easily obtained by an eavesdropper.

**Tag anonymity** Each tag updates IDS and private keys (k1, K2) after successful authentication, and this update process involves random numbers (n3, n'1, n'2). When the tag is interrogated again, a fresh IDS is backscattered. Additionally, all public submessages (A||B||C|| and D) are anonymized by the use of random numbers (n1, n2, n3, n'1). Tag anonymity is thus guaranteed, and location privacy of the tag owner is not compromised.

**Mutual Authentication and Data Integrity** The protocol provides mutual authentication. Only a legitimate reader possessing keys (k1, K2), can build a valid message A||B||C. Similarly, only a genuine tag can derive nonces n1, n2 from A||B||C, and then compute message D. Messages C and D, which involve the internal secret values (n3, n'1, k1*, k2*) and nonces (n1, n2), allow data integrity to be checked.

**Replay attacks** An eavesdropper could store all the messages exchanged in a protocol run. To impersonate the tag, he could replay message D. However, this response would be invalid as different nonces are employed in each session -this will frustrate this naive attack. Additionally, the attacker could pretend that the reader has not accomplished the updating phase in the previous session. In this scenario, the tag is identified by the old index-pseudonym and the attacker may forward the eavesdropped values of A||B||C. Even if this is successful, no secret information is disclosed and the internal state is unchanged in the genuine tag, so all these attacks are unsuccessful.

**Forward Security** Imagine that a tag is exposed one day, making public its secret information (ID, k1, K2). The attacker still cannot infer any information from previous sessions as two unknown nonces (n1, n2) and five internal secret values (n3, n'1, n'2, k1*, K2*) are involved in the message creation (mutual authentication phase).

Additionally, these internal values are employed in the updating phase. Consequently, past communications cannot be easily jeopardized.

**Updating Confirmation** The Gossamer protocol assumes that tags and readers share certain secret values. As these values are locally updated, synchronization is mandatory. Submessages C and D provide confirmation of the internal secret values (n3, n'1, k*1, k*2) and nonces (n1, n2). These values are employed in the updating stage. So the correct update of values IDS and keys (k1, k2) is implicitly ensured by submessages C and D. Unintentional transmission errors can happen in the received messages since a radio link is used. This is an extremely serious issue for message D, since it can result in a loss of synchronization. However, the tuple (IDS, k1, K2) is stored twice in the tag memory -once with the old values, the other with the potential next values. With this mechanism, even in the event that message D is incorrectly received, the tag and the reader can still authenticate with the old values. So the reader and the tag will be able to recover their synchronized state [2].

## 4   Passive Attacks against Gossamer

Attack 1: For this attack to work, the following condition has to be satisfied: Two nonces n1, n2 mod 96 = 0.

If *n1, n2 mod 96 = 0*
Then
*MixBits(0 mod 96,0 mod 96)=0 mod 96*                                                (3)
$C = ROT((ROT(n3+ K1*+\pi+n1', n3)+K2* \oplus n1', n2) \oplus n1'$
From **(3)**
*n1, n2, n3, n1' mod 96 all become ZEROs*
Hence
$C = K1* + \pi + K2*$
$K1*+K2*=C-\pi$                                                                        (4)
Similarly
$D = ROT((ROT(n2+K2*+ ID+n1', n2) +K1* +n1', n3) +n1'$
   $= K1*+ID+K2*$
$K1* + K2* = D - ID$                                                                   (5)
$IDS^{next} = ROT((ROT(n1'+K1*+IDS+n2',n1')+K2* \oplus n2', n3) \oplus n2'$
      $= K1* + IDS + K2*$
$K1* + K2* = IDS^{next} - IDS$                                                         (6)
From **(2)** and **(3)** $ID = D - C + \pi$                                            (7)
From **(3)** and **(4)** $ID = D - IDS^{next} + IDS$                                   (8)
From **(2)** and **(4)** $C-\pi = IDS^{next} - IDS$                                    (9)

While observing the external exchanged public messages, if two successive authentication sessions satisfy equation **(9)**, then ID value is determined.
Hence a passive attacker can find out the values of the secrete ID value.

Attack 2: If k1, K2 = 0 mod 96 Then:
$A = ROT((ROT(IDS+K1+\pi+n1, K2) +K1, K1)$
   $= IDS + \pi+n1$
Then $n1 = A - IDS – \pi$                                                  **n1 is known**
Similarly $B = ROT((ROT(IDS+K2+\pi+n2, K1) +K2, K2)$
          $= IDS + \pi+n1$
Then $n2 = B - IDS – \pi$                                                  **n2 is known**
$n3 = MixBits(n1, n2), n1' = MixBits(n3,n2)$                               **n1',n3 are known**
$K1* = ROT((ROT(n2+K1+\pi+n3, n2) +K2 \oplus n3, n1) \oplus n3$

$\quad\quad\quad =ROT((ROT(n2+\pi+n3,n2)\oplus n3,n1)\oplus n3$  $\quad\quad\quad$ *K1\* is known*

Similarly

$K2^* = ROT\ ((ROT\ (n1+K2+\pi+n3,\ n1) +K1+n3,\ n2)+n3$

$\quad\quad\ =ROT((ROT(n1+\pi+n3,\ n1)+n3\ n2)+n3$ $\quad\quad\quad$ *K2\* is known*

$C=ROT\ ((ROT\ (n3+K1^*+\pi+n1',\ n3)+K2^*\oplus n1',\ n2)\oplus n1'$

$\quad =ROT((ROT(n3+K1^*+\pi+n1',n3)+K2^*\oplus n1',n2)\oplus n1'$ $\quad\quad$ *C is known* **(10)**

$D=ROT\ ((ROT\ (n2+K2^* + ID+n1',\ n2)+K1^*+n1',\ n3)+n1'$

So $\ ID=ROT_{right}(ROT_{right}((D-n1'),n3)-K1^*-n1',n2)-n2-K2^*-n1'$ $\quad\quad\quad$ **(11)**

While observing the external exchanged public messages, if transmitted C = calculated C from **(10)**, then ID value can be determined from **(11)**.

And all secret values (k1, k2, n1, n2, n3, n1', n2', ID) will be known so $IDS^{next}$ can be determined also:

n2'= MIXBITS (n1 ', n3) $\quad\quad\quad\quad\quad$ *n2' is known*

$IDS^{next}=ROT((ROT(n1'+K1^*+IDS+n2',n1')+K2^*\oplus n2',n3)\oplus n2'$

Hence this attack can be done in one session.

*Note: The second attack is computationally infeasible since n1 is determined out of $2^{89}$ permutations.*

## 5    Proposed Solution

The paper proposes two modifications added to Gossamer protocol for making the protocol more secured against passive attacks:
- Make rotation functions dependent of nonces and keys values as shown in Figure.7 and Figure.8.
- MixBits function in Gossamer guarantees that in case both of its two inputs are zeros mod 96, its output will be zero mod 96. Hence modifying MixBits function as shown in Figure.9 to guarantee that in case of its two inputs is zeros mod 96; its output will not be zero mod 96 will increase the security of the protocol.

However the modifications of the proposed protocol do not affect computational, storage, or communication cost of Gossamer. It can be realistically implemented, even in low cost RFID tags.

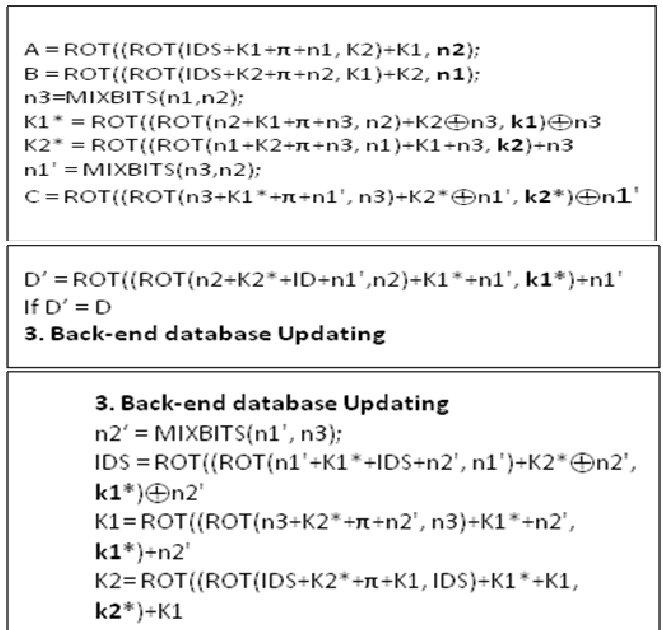

**Figure.7 Modified Gossamer at Reader**

```
Extract n1 from A, and n2 from B
n3=MIXBITS(n1, n2);
K1*=ROT((ROT(n2+K1+π+n3, n2)+K2⊕n3, k1)⊕n3
K2*=ROT((ROT(n1+K2+π+n3, n1)+K1+n3, k2)+n3
n1'=MIXBITS(n3, n2);
C'=ROT((ROT(n3+K1*+π+n1', n3)+K2*⊕n1', k2*)⊕n1'
If C' = C
D=ROT((ROT(n2+K2*+ID+n1',n2)+K1*+n1', k1*)+n1'

3. Tag Updating:
n2' = MIXBITS(n1',n3);
IDSold = IDS;
IDSnext= ROT((ROT(n1'+K1*+IDS+n2', n1')+K2*⊕n2', k1*)⊕n2'
K1old = K1;
K1next= ROT((ROT(n3+k2*+π+n2', n3)+K1*+n2', k1*)+n2'
K2old = K2;
K2next= ROT((ROT(IDSnext+K2*+π+K1next, IDSnext)+K1*+K1next, k2*)+K1next
```

**Figure.8 Modified Gossamer at tag**

```
Z = MixBits(X, Y)
{   Z = X;
    for (i=0; i< 32; i++)
    {Z = (Z+i) + Z + Z + Y ;}
}
```

**Figure.9 Modified MixBits**

# 6    Analysis of the Proposed Solution

Analyzing the security and performance of the proposed protocol shows that the added modifications increase security level of Gossamer and prevent eavesdropping on public messages between reader and tag.
Considering the former first attack scenario, ID cannot be determined as follows:
*If n1, n2 mod 96 =0*
*Then*
*A = ROT (IDS+K1+π, K2) +K1*
*B = ROT (IDS+K2+π, K1) +K2*
*K1* = ROT (K1+π+n3+ K2⊕n3, k1) ⊕n3*
*K2* = ROT (K2+π+K1+2n3, K2) +n3*
*C = ROT ((ROT (n3+K1*+π+n1', n3) +K2*⊕n1', K2*) ⊕n1'*
*D = ROT ((ROT (n2+K2*+ID+n1', n2) +K1*+n1', k1*) +n1'*
*   = ROT (K2*+ID+n1'+K1*+n1', k1*) +n1'.*
Then from these set of equations ID cannot be determined.

Considering the former second attack scenario, ID cannot be determined as follows:
*If k1, k2 mod 96 =0*
*A = ROT ((ROT (IDS+K1+π+n1, K2) +K1, n2)*
*B = ROT ((ROT (IDS+K2+π+n2, K1) +K2, n1)*
*Then:-*
*A = ROT (IDS+π+n1, n2)*
*B = ROT (IDS+π+n2, n1)*
*K1*=ROT (n2+π+n3, n2)*
*K2*=ROT (n1+π+n3, n1) +n3+n3*
*C=ROT ((ROT (n3+K1*+π+n1', n3) +K2*⊕n1', K2*) ⊕n1*
*D=ROT ((ROT (n2+K2*+ID+n1', n2) +K1*+n1', k1*) +n1'.*
*  =ROT (K2*+ID+n1'+K1*+n1', k1*) +n1'.*
Then from these set of equations ID cannot determined.

Even if we assume that *k1, K2, n1, n2 mod 96 =0* at the same time, ID cannot be determined as follows:
A = IDS + π
B = IDS + π
As n3 = MixBits(n1,n2)    n3 ≠ 0 mod 96
K1* = π + n3
K2* = π + 3n3
C = ROT ((ROT (n3+K1*+π+n1', n3) +K2*⊕n1', K2*) ⊕n1'
D = ROT ((ROT (n2+K2*+ID+n1', n2) +K1*+n1', k1*) +n1'
ROT (K2*+ID+n1'+ K1*+n1', k1*) +n1'.
Then from these set of equations ID cannot determined.

# 7    Conclusion

This paper has proposed modifications to Gossamer mutual authentication protocol for low cost RFID tags. The proposed protocol prevents passive attacks as active attacks are discounted when designing a protocol to meet the requirements of low cost RFID tags. The analysis of the proposed protocol shows that the added modifications increase security level of Gossamer and prevent eavesdropping on public messages between reader and tag. However the modifications of the proposed protocol do not affect computational, storage, or communication cost of Gossamer. It can be realistically implemented, even in low cost RFID tags.

## Authors

**Eslam Gamal Ahmed**

Teacher assistant in computer systems department in faculty of computer and information sciences, Ain shams university. Graduated at June 2006 with a degree excellent with honor. Currently I am working in my master degree titled "Developing Lightweight Cryptography for RFID ".Fields of interest are computer and networks security, RFID systems and computer architecture.

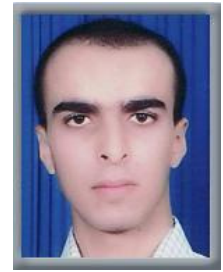

**Dr. Eman Shaaban**

Lecturer in computer systems department in faculty of computer and information sciences, Ain shams university. Fields of interest are RFID systems, embedded systems, computer and networks security, logic design and computer architecture.

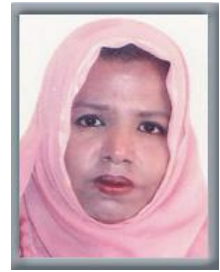

**Prof.Mohamed Hashem**

Department head of information systems department in faculty of computer and information sciences, Ain shams university. Fields of interest are computer networks, Ad-hoc and wireless networks, Qos Routing of wired and wireless networks, Modeling and simulation of computer networks, VANET and computer and network security.

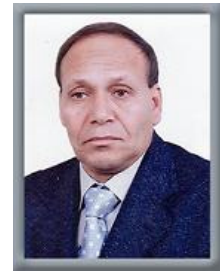